\documentclass[preprint,showpacs,superscriptaddress,aps,a4paper]{revtex4}
\usepackage{pstricks,pst-node,pst-text,pst-3d,graphpap,pst-plot}
\usepackage{dcolumn}
\usepackage{amsmath}
\usepackage{graphicx}
\usepackage{latexsym}
\usepackage{amsfonts}
\usepackage{amssymb}
\DeclareGraphicsExtensions{.pdf,.gif,.jpg}

\newcommand{\be}{\begin{equation}}
\newcommand{\ee}{\end{equation}}
\newcommand{\beq}{\begin{eqnarray}}
\newcommand{\eeq}{\end{eqnarray}}

\begin{document}

\def\bbe{\mbox{\boldmath $e$}}
\def\bbf{\mbox{\boldmath $f$}}
\def\bg{\mbox{\boldmath $g$}}
\def\bh{\mbox{\boldmath $h$}}
\def\bj{\mbox{\boldmath $j$}}
\def\bq{\mbox{\boldmath $q$}}
\def\bp{\mbox{\boldmath $p$}}
\def\br{\mbox{\boldmath $r$}}
\def\bz{\mbox{\boldmath $z$}}

\def\bfzero{\mbox{\boldmath $0$}}
\def\bfone{\mbox{\boldmath $1$}}

\def\dr{{\rm d}}

\def\tb{\bar{t}}
\def\zb{\bar{z}}

\def\tgb{\bar{\tau}}

\def\bC{\mbox{\boldmath $C$}}
\def\bG{\mbox{\boldmath $G$}}
\def\bH{\mbox{\boldmath $H$}}
\def\bK{\mbox{\boldmath $K$}}
\def\bM{\mbox{\boldmath $M$}}
\def\bN{\mbox{\boldmath $N$}}
\def\bO{\mbox{\boldmath $O$}}
\def\bQ{\mbox{\boldmath $Q$}}
\def\bR{\mbox{\boldmath $R$}}
\def\bS{\mbox{\boldmath $S$}}
\def\bT{\mbox{\boldmath $T$}}
\def\bU{\mbox{\boldmath $U$}}
\def\bV{\mbox{\boldmath $V$}}
\def\bZ{\mbox{\boldmath $Z$}}

\def\hH{\mbox{$\hat{H}$}}

\def\bcalS{\mbox{\boldmath $\mathcal{S}$}}
\def\bcalG{\mbox{\boldmath $\mathcal{G}$}}
\def\bcalE{\mbox{\boldmath $\mathcal{E}$}}

\def\bgG{\mbox{\boldmath $\Gamma$}}
\def\bgL{\mbox{\boldmath $\Lambda$}}
\def\bgS{\mbox{\boldmath $\Sigma$}}

\def\bgr{\mbox{\boldmath $\rho$}}
\def\bgs{\mbox{\boldmath $\sigma$}}

\def\a{\alpha}
\def\b{\beta}
\def\g{\gamma}
\def\G{\Gamma}
\def\d{\delta}
\def\D{\Delta}
\def\e{\epsilon}
\def\ve{\varepsilon}
\def\z{\zeta}
\def\h{\eta}
\def\th{\theta}
\def\k{\kappa}
\def\l{\lambda}
\def\L{\Lambda}
\def\m{\mu}
\def\n{\nu}
\def\x{\xi}
\def\X{\Xi}
\def\p{\pi}
\def\P{\Pi}
\def\r{\rho}
\def\s{\sigma}
\def\S{\Sigma}
\def\t{\tau}
\def\f{\phi}
\def\vf{\varphi}
\def\F{\Phi}
\def\c{\chi}
\def\w{\omega}
\def\W{\Omega}
\def\Q{\Psi}
\def\q{\psi}

\def\ua{\uparrow}
\def\da{\downarrow}
\def\de{\partial}
\def\inf{\infty}
\def\ra{\rightarrow}
\def\bra{\langle}
\def\ket{\rangle}
\def\grad{\mbox{\boldmath $\nabla$}}
\def\Tr{{\rm Tr}}
\def\hc{{\rm h.c.}}

\title{Circulating currents and magnetic moments in quantum rings}

\author{Michele Cini}
\affiliation{Dipartimento di Fisica, Universit\`a di Roma Tor
Vergata, Via della Ricerca Scientifica 1, 00133 Rome, Italy}
\affiliation{Istituto Nazionale
di Fisica Nucleare, Laboratori Nazionali di Frascati, Via E. Fermi 40, 00044 Frascati, Italy}

\author{Enrico Perfetto}
\affiliation{Unit\`a CNISM, Universit\`a di Roma Tor Vergata,
Via della Ricerca Scientifica 1, 00133 Rome, Italy}

\author{Gianluca Stefanucci}
\affiliation{Dipartimento di Fisica, Universit\`a di Roma Tor
Vergata, Via della Ricerca Scientifica 1, 00133 Rome, Italy}
\affiliation{Istituto Nazionale
di Fisica Nucleare, Laboratori Nazionali di Frascati, Via E. Fermi 40, 00044 Frascati, Italy}
\affiliation{European Theoretical Spectroscopy Facility (ETSF)}


\begin{abstract}
In circuits containing closed loops the operator for the current is
determined by charge conservation up to an arbitrary divergenceless
current.
In this work we propose a formula to calculate the magnetically
active circulating current $I_{\rm ring}$ flowing along a quantum
ring connected to biased leads. By {\em gedanken experiments} we argue
that $I_{\rm ring}$ can be obtained from the response of the
gran-canonical energy of the ring to an external magnetic flux.
The results agree with those of the conventional approach in the case of
isolated rings. However, for connected rings $I_{\rm ring}$ cannot be
obtained as a linear combination of bond currents.

\end{abstract}


\maketitle
\date{\today}

In this Letter we show that the theory of quantum
transport \cite{l.1957,caroli,cini80,b.1986,hj.2008,topics} must be
extended when dealing with circuits containing closed loops
in order to calculate the magnetic moment which couples to an external
magnetic field. In the quantum theory of transport the operator\cite{caroli}\cite{units}
\be
\hat{J}_{mn}=-i( t_{mn}c^{\dag}_{m}c_{n}-t_{nm}c^{\dag}_{n}c_{m})
\label{scossa}
\ee
is interpreted as the electron current operator
between sites $m$ and $n$ connected by a bond
with hopping integral $t_{mn}$. Such
interpretation naturally stems for the continuity equation
\be
\frac{d}{dt}\hat{n}_{m}=\sum_{n}\hat{J}_{mn}
\ee
in which the change in density $\hat{n}_{m}$ on site $m$ is seen as
the sum of the currents flowing from site $m$ to all connected sites
$n$. In a similar way one obtains the formula for the current density in a
continuum system.
We will  call $\hat{J}_{mn}$ the {\em bond current} operator since
it depends on the operators straddling a  bond.
In the case of a ring, however, and in general for circuits
containing loops,
the continuity alone cannot uniquely fix the current since one remains
free to add a  divergence-less component.
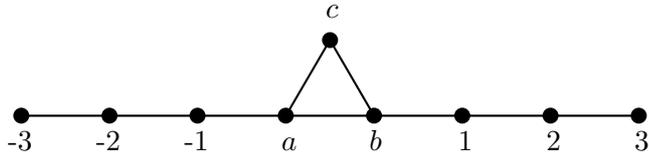
\begin{figure}[t]
\begin{center}
    \psset{unit=.58cm}
    \begin{pspicture}(-8,-0.4)(5.657,1.414)
    \psdots[dotsize=6pt,linecolor=black](-8,0)
    \psdots[dotsize=6pt,linecolor=black](-6,0)
    \psdots[dotsize=6pt,linecolor=black](-4,0)
    \psdots[dotsize=6pt,linecolor=black](-2,0)
    \psdots[dotsize=6pt,linecolor=black](-1,1.73)
    \psdots[dotsize=6pt,linecolor=black](0,0)
    \psdots[dotsize=6pt,linecolor=black](2,0)
    \psdots[dotsize=6pt,linecolor=black](4,0)
    \psdots[dotsize=6pt,linecolor=black](6,0)
    \psline(-8,0)(6,0)
    \psline(-2,0)(-1,1.73)
    \psline(0,0)(-1,1.73)
    \put(-8.3,-0.8){-3}
    \put(-6.3,-0.8){-2}
    \put(-4.3,-0.8){-1}
    \put(-2.1,-0.8){$a$}
    \put(-0.1,-0.8){$b$}
    \put(-1.1,2.23){$c$}
    \put(1.9,-0.8){1}
    \put(3.9,-0.8){2}
    \put(5.9,-0.8){3}
    \end{pspicture}
\end{center}
\caption{\footnotesize  Sketch of a ring with three sites in contact
with two semiinfinite one-dimensional leads. The site of the ring are labeled
with Roman letters, while the site of the left/right lead are
labeled with negative/positive integers.}
\label{triangolo}
\end{figure}
Such circuits are not merely accademic. The experimental realization
of mesoscopic metallic rings\cite{expring} has prompted an extraordinary research activity
on the quantum behavior of electrons and fundamental
paradimgs like Aharonov-Bohm oscillations\cite{guinea} and persistent currents \cite{maiti} are
currently under intense investigation. Recent progresses in
connecting aromatic molecules to metallic leads have brought
the ring-like topology into the nano-world as well \cite{molel}.

Below we specialize the discussion to tight-binding rings
connected to biased leads for the sake of definiteness. The continuum counterpart
is affected by the same ambiguity and deserves a similar discussion.
These systems have been mainly considered to study the quantum
interference pattern of the total current
\cite{ycp.2003,matthias,pf.2008} and of the
ring bond-currents\cite{jd.1995,noi}. Nevertheless, scarce attention has so far been given to the
calculation of the ring magnetic moment.
The  current pattern along the ring
is a superposition of a circulating current $I_{\rm ring}$ which is
magnetically active and a  laminar one. How to calculate
$I_{\rm ring}$ is the main contribution of this Letter.

Let us consider the tight-binding model of Fig. \ref{triangolo}
described by the Hamiltonian $\hat{H}$
\be
\hat{H}=t\sum_{\bra m,n\ket }c^{\dag}_{m}c_{n},
\ee
where $\bra \ldots\ket$ denotes nearest neighbor sites.
Due to the ambiguity discussed above it is not granted that the
physical current flowing through bonds $a-b$ and $a-c$
is the same as the expectation value
$J_{ab}$ and $J_{ac}$ of the bond current operator in Eq. (\ref{scossa}),
so we must ask how the
current pattern could be measured.
In a macroscopic ring connected to
leads, one can get the current in each wire
by using an amperometer or by  exploring
the magnetic field around each branch of the circuit and performing
the line integral.
However, for a quantum ring this cannot be done and in principle
the  coherence between alternative paths (that electrons explore
$a-c-b$ and $a-b$ simultaneously) defies every bond-related
definition of the circulating  current.

In the case of an isolated ring with Hamiltonian $\hat{H}_{\rm
ring}(\f)$,  the current $I^{\rm isolated}_{\rm ring}$ that can
couple to a magnetic flux $\phi$ and generate the ring
magnetic moment is\cite{kohn,agnese} \be I^{\rm isolated}_{\rm ring}=c \frac{d\bra \hat{H}_{\rm
ring}(\f)\ket}{d\phi}.\label{acgauge}
\ee    Here, the
magnetic flux is  $\f=\a_{ab}+\a_{bc}+\a_{ca}$ where the
$\a$'s are the phases of the hopping integrals
$t_{mn}=|t_{mn}|e^{i\frac{\a_{mn}}{c}}$, in accordance with the Peierls
prescription. Experimentally, one could get $I^{\rm isolated}_{\rm ring}$
by measuring the torque acting on the ring in a magnetic field.
There is no ambiguity, since  $I^{\rm isolated}_{\rm ring}\equiv J_{ab} = J_{ca}$
is the only physical current.

In this work we   define $I_{\rm ring}$ for the connected ring by  a proper  magnetic
measurement to be performed \textit{in situ} on the ring itself.
We shall see that  the Hamiltonian contains enough
information to compute $I_{\rm ring},$ since   the coupling to
an external field via the Peierls prescription encodes the necessary
information. To this end, we must introduce local force measurements and illustrate the idea by electric and magnetic thought experiments in parallel, since the two cases illuminate each other. We wish  to show that in both cases we need a local probe and a local readout of the result.

\begin{figure}[htbp]
\includegraphics*[width=.4\textwidth]{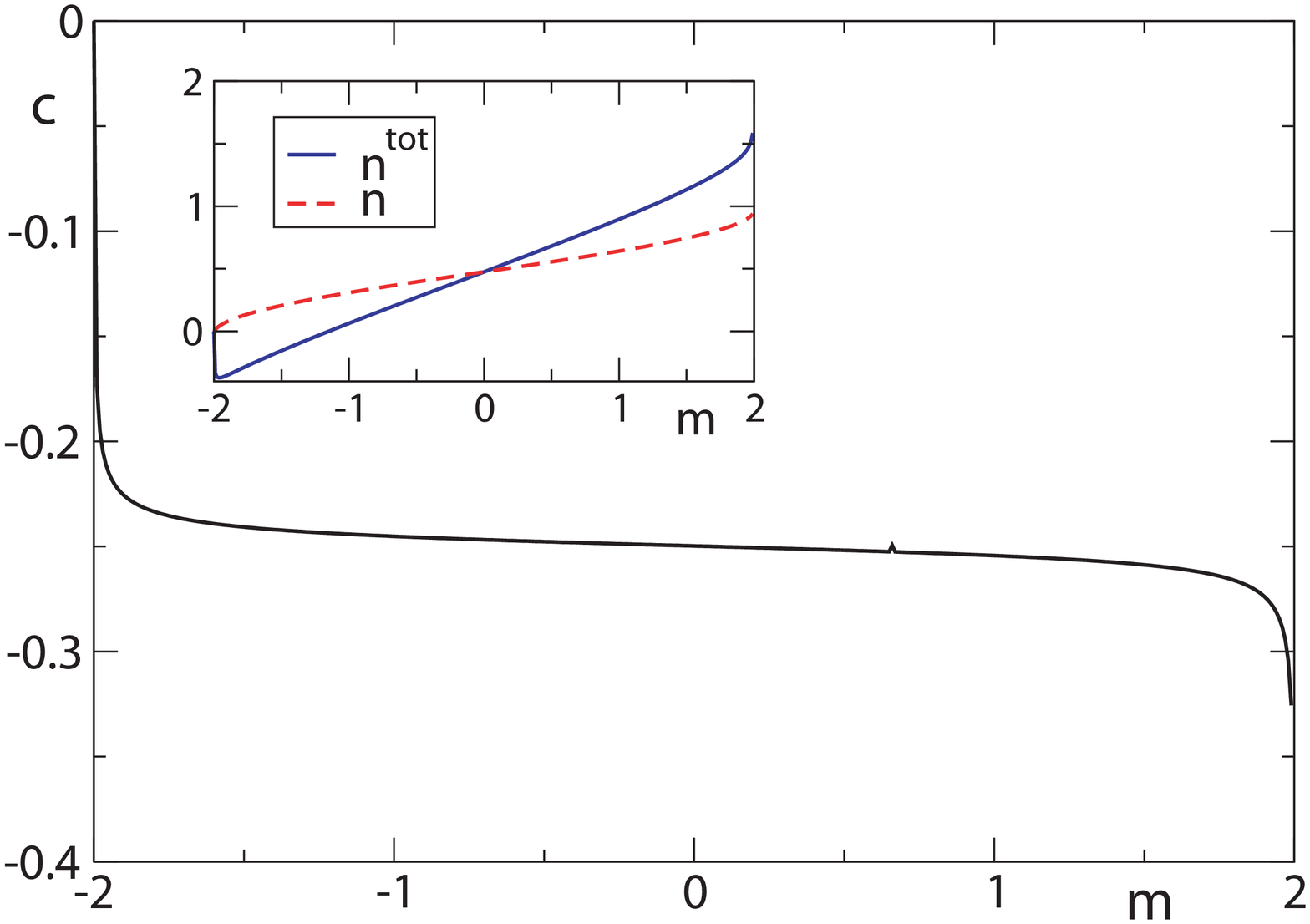}
\caption{(Color online) Local response
$\chi_{m}$ for a one-dimensional tight-binding chain
as a function of the chemical potential $\m$ for $\ve_{m}=0$.
The inset shows the ground state local density
$n_{m}$ (red, dashed) and
the total response density $n^{\rm tot}_{m}$ (blue, solid) as in Eq. (\ref{densresp}).
}
\label{density}
\end{figure}

{\em Electrostatic experiment}.
  Suppose a macroscopic  circuit is prepared in  the  eigenstate $|\Q\ket$ of the Hamiltonian $\hat{H}$. One can gain information  about the charge and polarisability at some site $m$ of the system  by an intensive measurement, e.g. by measuring a force. As a probe, one
 could use a tiny condenser to set a  weak electric field
of strength $D$
directed, say, along the $\hat{x}$ axis, right at site
$m$.
The on-site  field shifts the atom by $x$ and changes the site energy  accordingly, $\ve_{m}\ra\ve_{m}(x)=\ve_{m}+Dx$, while the ground state  becomes
$|\Q(x)\ket$ with  $|\Q(0)\ket=|\Q\ket$. Now, we must decide precisely which  force to measure as a response to the local probe.
If the whole circuit could be treated like a rigid body, one could measure
 $F=-\frac{d E}{dx}|_{x=0}$
where $E(x)=\langle \Q(x) |\hH(x)|\Q(x)\rangle$ is the total energy; $F$ is {\em total} force on the system.
By exploiting the Hellmann-Feynman theorem one finds that the  observable is
$-F/D=n_{m}=\bra\Q|\hat{n}_{m}|\Q\ket$, i.e., the
average density on site $m$, which  is often obtainable by simpler means.
 However, as the probing
electric field is local,  a local readout of the experiment is desirable.
The on-site force on the atom can be measured using, e.g., an atomic force microscope
and can be expressed in terms of the {\em local} energy as
\be
F=-\frac{d E_{\rm at}}{dx}|_{x=0},
\ee
where
\be
E_{\rm at}(x)=\langle \Q(x)|(\ve_{m}(x)-\mu) \hat{n}_{m}|\Q(x)\rangle
\ee
is the grand-canonical energy of site $m$ with chemical potential $\m$.
The external circuit works as a reservoir for particles and heat. The
use of the grand-canonical formalism ensures the gauge invariance of the
theory versus shifts of the energy origin. The  local force measurement yields $F=-n^{\rm tot}_{m}D$ where
$n^{\rm tot}_{m}=\langle \Q| \hat{n}_{m}|\Q \rangle+\d n_{m}$ with
\be
\d n_{m}=\frac{\ve_{m}-\mu}{D}\frac{d}{d x}\left.\langle
\Q(x)| \hat{n}_{m}|\Q( x) \rangle\right|_{x=0}.
\label{densresp}
\ee
The   extra  contribution $\d n_{m}$  is interesting  since
$\d n_{m}=(\ve_{m}-\mu)\chi_{m}$
where $\chi_{m}=\frac{1}{D}\frac{d}{d x}\left.\langle
\Q(x)| \hat{n}_{m}|\Q( x) \rangle\right|_{x=0}$
is the ratio of the polarization charge $d \langle
\Q(x)| \hat{n}_{m}|\Q( x) \rangle $ to the external potential $D dx$ induced by a small shift of the atom in the field. Thus  $\chi_{m}$  brings
information on the local dielectric response, while
the factor $\ve_{m}-\mu$
accounts for the work done to bring charge  from infinity to
site $m$. For example, if the overall charge on the atom is negative, the field will shift it towards positive potentials, hence the site will be more attractive for electrons and $Dx<0;$ thus we may predict that $d \langle
\Q(x)| \hat{n}_{m}|\Q( x) \rangle >0$ and $\chi_{m}<0.$ Then if $\ve_{m}<\m,$ that is, the level is more than half filled, we may expect a positive $\d n_{m},$ and a further increase of the electron population, while less than half filled levels will tend to be emptied.
In Fig. \ref{density} we display the ground state response
$\chi_{m}$ as well as the densities $n_{m}$ and $n_{m}^{\rm tot}$
for a one-dimensional tight-binding chain with nearest neighbor hopping
$t$ and zero on-site energy, $\ve_{m}=0$, as a function of the
chemical potential $\m$.  Note  that at
high filling the local response of the system does not vanish due
to  a split-off state at energy larger than 2.

\begin{figure}
\begin{center}
    \psset{unit=.58cm}
    \begin{pspicture}(-8,-9.4)(5.657,3)
    \psdots[dotsize=6pt,linecolor=black](-8,0)
    \psdots[dotsize=6pt,linecolor=black](-6,0)
    \psdots[dotsize=6pt,linecolor=black](-4,0)
    \psdots[dotsize=6pt,linecolor=black](-2,0)
    \psdots[dotsize=6pt,linecolor=black](-1,1.73)
    \psdots[dotsize=6pt,linecolor=black](0,0)
    \psdots[dotsize=6pt,linecolor=black](2,0)
    \psdots[dotsize=6pt,linecolor=black](4,0)
    \psdots[dotsize=6pt,linecolor=black](6,0)
    \psline(-8,0)(6,0)
    \psline(-2,0)(-1,1.73)
    \psline(0,0)(-1,1.73)
    \put(-8.3,-0.8){-3}
    \put(-6.3,-0.8){-2}
    \put(-4.3,-0.8){-1}
    \put(-2.1,-0.8){$a$}
    \put(-0.1,-0.8){$b$}
    \put(-1.1,2.23){$c$}
    \put(1.9,-0.8){1}
    \put(3.9,-0.8){2}
    \put(5.9,-0.8){3}
    \put(-1.1,-0.5){$\f$}
    \put(1.9,1.){choice c1: $\q_{k}$}

    \psdots[dotsize=6pt,linecolor=black](-8,-4.5)
    \psdots[dotsize=6pt,linecolor=black](-6,-4.5)
    \psdots[dotsize=6pt,linecolor=black](-4,-4.5)
    \psdots[dotsize=6pt,linecolor=black](-2,-4.5)
    \psdots[dotsize=6pt,linecolor=black](-1,-2.77)
    \psdots[dotsize=6pt,linecolor=black](0,-4.5)
    \psdots[dotsize=6pt,linecolor=black](2,-4.5)
    \psdots[dotsize=6pt,linecolor=black](4,-4.5)
    \psdots[dotsize=6pt,linecolor=black](6,-4.5)
    \psline(-8,-4.5)(6,-4.5)
    \psline(-2,-4.5)(-1,-2.77)
    \psline(0,-4.5)(-1,-2.77)
    \put(-8.3,-5.3){-3}
    \put(-6.3,-5.3){-2}
    \put(-4.3,-5.3){-1}
    \put(-2.1,-5.3){$a$}
    \put(-0.1,-5.3){$b$}
    \put(-1.1,-2.27){$c$}
    \put(1.9,-5.3){1}
    \put(3.9,-5.3){2}
    \put(5.9,-5.3){3}
    \put(-2,-3.7){$\f$}
    \put(1.9,-3.5){choice c2: $\tilde{\q}_{k}$}

    \psdots[dotsize=6pt,linecolor=black](-8,-9)
    \psdots[dotsize=6pt,linecolor=black](-6,-9)
    \psdots[dotsize=6pt,linecolor=black](-4,-9)
    \psdots[dotsize=6pt,linecolor=black](-2,-9)
    \psdots[dotsize=6pt,linecolor=black](-1,-7.27)
    \psdots[dotsize=6pt,linecolor=black](0,-9)
    \psdots[dotsize=6pt,linecolor=black](2,-9)
    \psdots[dotsize=6pt,linecolor=black](4,-9)
    \psdots[dotsize=6pt,linecolor=black](6,-9)
    \psline(-8,-9)(6,-9)
    \psline(-2,-9)(-1,-7.27)
    \psline(0,-9)(-1,-7.27)
    \put(-8.3,-9.8){-3}
    \put(-6.3,-9.8){-2}
    \put(-4.3,-9.8){-1}
    \put(-2.1,-9.8){$a$}
    \put(-0.1,-9.8){$b$}
    \put(-1.1,-6.77){$c$}
    \put(1.9,-9.8){1}
    \put(3.9,-9.8){2}
    \put(5.9,-9.8){3}
    \put(-1.1,-9.5){$\f$}
    \put(2.6,-9.5){$-\f$}
    \put(1.9,-8){choice c2$'$: $\tilde{\q}'_{k}$}
    \end{pspicture}
\end{center}
\caption{ Configuration of the Peierls phases along the
bonds of the system. For choice c1 the phase is localized on the bond
$a-b$ while for choice c2 the phase is along the bond $a-c$.
Starting from c2, the multiplication of the fermion operators on sites $c$, $b$, and $-1$ by
$e^{i\f}$ is a gauge transformation which corresponds to the
configuration c2$'$.}
\label{gauge}
\end{figure}
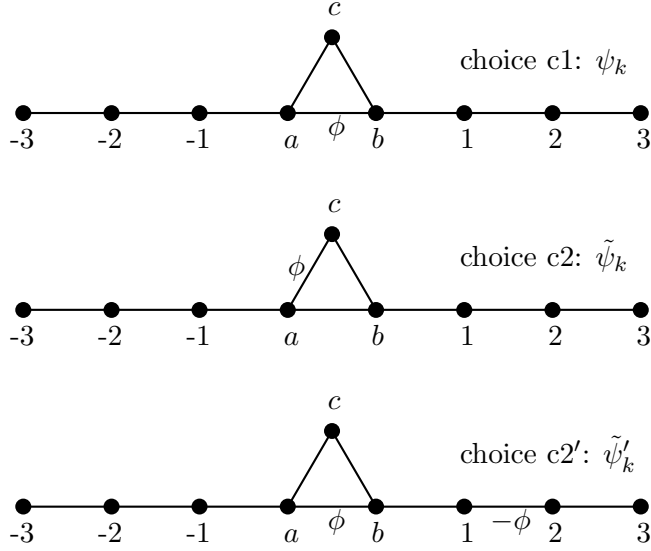

{\em Magnetic experiment}. We  develop the magnetic case by a
{\em gedanken} experiment as far as possible in parallel with  the
above, aiming at a definition of the ring current that corresponds
to Eq. (\ref{densresp}).  While the leads are biased and the current flows, we
switch a weak uniform magnetic field of strength $B$ forming an
angle $\theta$ with respect to the normal of the ring. We denote
with $\f$ the magnetic flux through the ring and with
$|\Q(\f)\ket$ the current carrying state of the system after all
transient effects have disappeared. The derivative of the total
energy $E(\f)=\langle\Q(\f)| \hat{H}(\f)|\Q(\f)  \rangle$ would
give the torque acting on the whole system. As in the
electrostatic experiment, the wave function is modified everywhere
even though the magnetic perturbation is localized and hence
$\langle\Q(\phi)| \hH(\f)-\hH_{\rm ring}(\f)|\Q(\phi)\rangle $
depends on the flux and the external circuit experiences a torque
as well. A further problem in using the variation of the total
energy is related to the choices of the Peierls phases along the
ring. The definition of the local torque should only depend on the
magnetic flux $\f=\a_{ab}+\a_{bc}+\a_{ca}$ through the ring. However, one can
easily realize that the choice $\a_{ab}=\f$ and
$\a_{bc}=\a_{ca}=0$ (c1) and the choice $\a_{ab}=\a_{bc}=0$ and
$\a_{ca}=\f$ (c2), see Fig. \ref{gauge}, are {\em not} related via a
gauge transformation and hence lead to  different derivatives of the total energy with
respect to $\f$. The asymptotic current-carrying eigenstate\cite{noi} of $H$
can be used in both cases to invoke the   Hellmann-Feynman theorem.
The choice c1 then leads to the bond current $J_{ab}$ while the
choice c2 to  $J_{ca}$. The use of the total energy
to calculate the torque experienced by the ring is therefore
ambiguous. Next,  we show that   a local
measurement on the ring is much more rewarding.

The ring interacts with the magnetic field and its energy relative to
the chemical potential $\m$ is
\beq
E_{\rm ring}(\f)&=&\bra\Q(\f)|\hH_{\rm ring}(\f)-\m\hat{N}_{\rm ring}|\Q(\phi)\ket
\nonumber \\
&=&-M B \cos\theta=-M \frac{\phi}{S},
\label{ering}
\eeq
where $\hat{N}_{\rm ring}$ is the number operator of the ring,
$S$ is the ring surface and $M$ is the ring magnetic moment.
We have no information about $S$,  however, we may say that
\be
\frac{1}{c}I_{\rm ring}=-\frac{M}{S}=\frac{d E_{\rm ring}}{d \phi}|_{\phi=0}.
\label{iring2}
\ee
The definition of the ring current does not suffer from the ambiguity
originating from different possible choices of the Peierls phases.
This follows from the fact that the operator $\hH_{\rm
ring}(\f)-\m\hat{N}_{\rm ring}$ is a local operator and hence its average
only depends on the projection onto the ring of the single particle
states $\{\q_{k}\}$ forming
the Slater determinant $|\Q(\f)\ket$. Let us discuss this crucial
point in more detail.
We consider the choice c1 and let $\q_{k}(m)$ be the amplitude on site $m$
of the $k$-th one-particle eigenstate of the total Hamiltonian, see
Fig. (\ref{gauge}).
Similarly we denote with $\tilde{\q}_{k}(m)$ the one-particle
eigenstate corresponding to the choice c2.
The choice c2 can be transformed in a gauge equivalent phase
configuration (choice c2$'$) in which the Peierls phases along the ring are the same as
those in c1 but in the right lead there is a bond, e.g., the bond
$1-2$, that acquires a phase $-\f$. The eigenstate $\tilde{\q}_{k}(m)$
of choice c2 is transformed in the eigenstate $\tilde{\q}'_{k}(m)$
accordingly. It is straightforward to realize that
$\tilde{\q}'_{k}(m)=e^{i\frac{\f}{c}}\q_{k}(m)$ for $m>1$ while
$\tilde{\q}'_{k}(m)=\q_{k}(m)$ otherwise. As a consequence the
average of the local operator $\hH_{\rm
ring}(\f)-\m\hat{N}_{\rm ring}$ over the gauge inequivalent
configurations c1 and c2 does not change. Such result is independent
of the choice of the Peiers phases along the ring provided that
$\a_{ab}+\a_{bc}+\a_{ca}=\f$.

{\em Results and discussion}. To calculate the ring current from Eq.
(\ref{iring2}) we use an embedding technique.
We consider the system of Fig. \ref{triangolo} with zero on site
energies everywhere and hopping $|t|$ between the sites connected by
a link. Let $U_{L/R}$ be the
bias applied to the left/right lead and $h$ be the
matrix of the one-body operator $\hH_{\rm
ring}(\f)=\sum_{ij}h_{ij}(\f)c^{\dag}_{i}c_{j}$. Then the average
$E_{\rm ring}(\f)$ in Eq. (\ref{ering})
can be expressed in terms of the lesser Green's function $G^{<}$ as
\be
E_{\rm ring}(\f)=
-i\int\frac{d\w}{2\p}\Tr_{\rm ring}\left[(h-\m)G^{<}(\w)\right],
\label{ermb}
\ee
where the trace is taken over the sites of the ring. The matrix
$G^{<}(\w)$ is the Fourier transform of the  lesser Green's function
$G^{<}(t,t')$ for times $t,t'\ra\inf$ and can be written as
$G^{<}(\w)=G^{\rm R}(\w)[\S^{<}_{L}(\w)+\S^{<}_{R}(\w)]G^{\rm A}(\w)$.
The retarded/advanced Green's function projected onto
the ring is $G^{\rm R/A}(\w)=[\w-h-\S^{\rm R/A}(\w)]^{-1}$ with
$\S^{\rm R/A}=\S^{\rm R/A}_{L}+\S^{\rm R/A}_{R}$ the
retarded/advanced embedding self-energy of the left and right leads.
Using the fluctuation-dissipation theorem for lead $\a=L,R$ one
obtains for the
lesser embedding self-energy
$\S^{<}_{\a}(\w)=-2if(\w-U_{\a}){\rm Im}[\S^{\rm R}_{\a}(\w)]$, where
$f(\w)$ is the Fermi distribution function at chemical potential $\m$.
The retarded and advanced components are related as $\S^{\rm
R}_{\a}=[\S^{\rm A}_{\a}]^{\dag}$ with $\a=L,R$.
For one-dimensional tight-binding leads with nearest neighbor hopping $t$
the self-energies have only one
nonvanishing matrix element, namely
$[\S^{\rm R}_{L}(\w)]_{ij}=\d_{ia}\d_{ja}\s(\w-U_{L})$ and
$[\S^{\rm R}_{R}(\w)]_{ij}=\d_{ib}\d_{jb}\s(\w-U_{R})$. The function
$\s(\w)$ can be easily calculated and reads
\be
\s(\w)=\frac{1}{2}\left(
(\w+i\eta)-\frac{(\w+i\eta)+2t}{\sqrt{1+\frac{4t}{(\w+i\eta)-2t}}}
\right).
\ee
The ring current in Eq.(\ref{iring2}) is obtained by taking
the flux derivative of $E_{\rm ring}(\f)$ in Eq. (\ref{ermb})
in $\f=0$. The flux derivative of the one-body matrix hamiltonian $h$
yields a linear combination of the one-body matrix bond currents
with coefficients $\a_{ab}/\f$, $\a_{bc}/\f$ and $\a_{ca}/\f$.
This term alone would then be dependent on the Peierls phase configuration.
The independence of $I_{\rm ring}$ from the phase configuration is
restored by adding the flux derivative of $G^{<}(\w)$ which reads
\be
\frac{d}{d\f}G^{<}=G^{\rm R}\left(
\frac{d h}{\d\f}G^{\rm R}\S^{<}+\S^{<}G^{\rm A}\frac{d h}{d \f}
\right)G^{\rm A}.
\ee
Thus $I_{\rm ring}$ can be expressed in terms of
$\frac{d h}{d\f}|_{\f=0}$ and Green's functions at $\f=0$. We wish to
emphasize that for $\f=0$ the ring current has no diamagnetic
contribution.

\begin{figure}[htbp]
\includegraphics*[width=.47\textwidth]{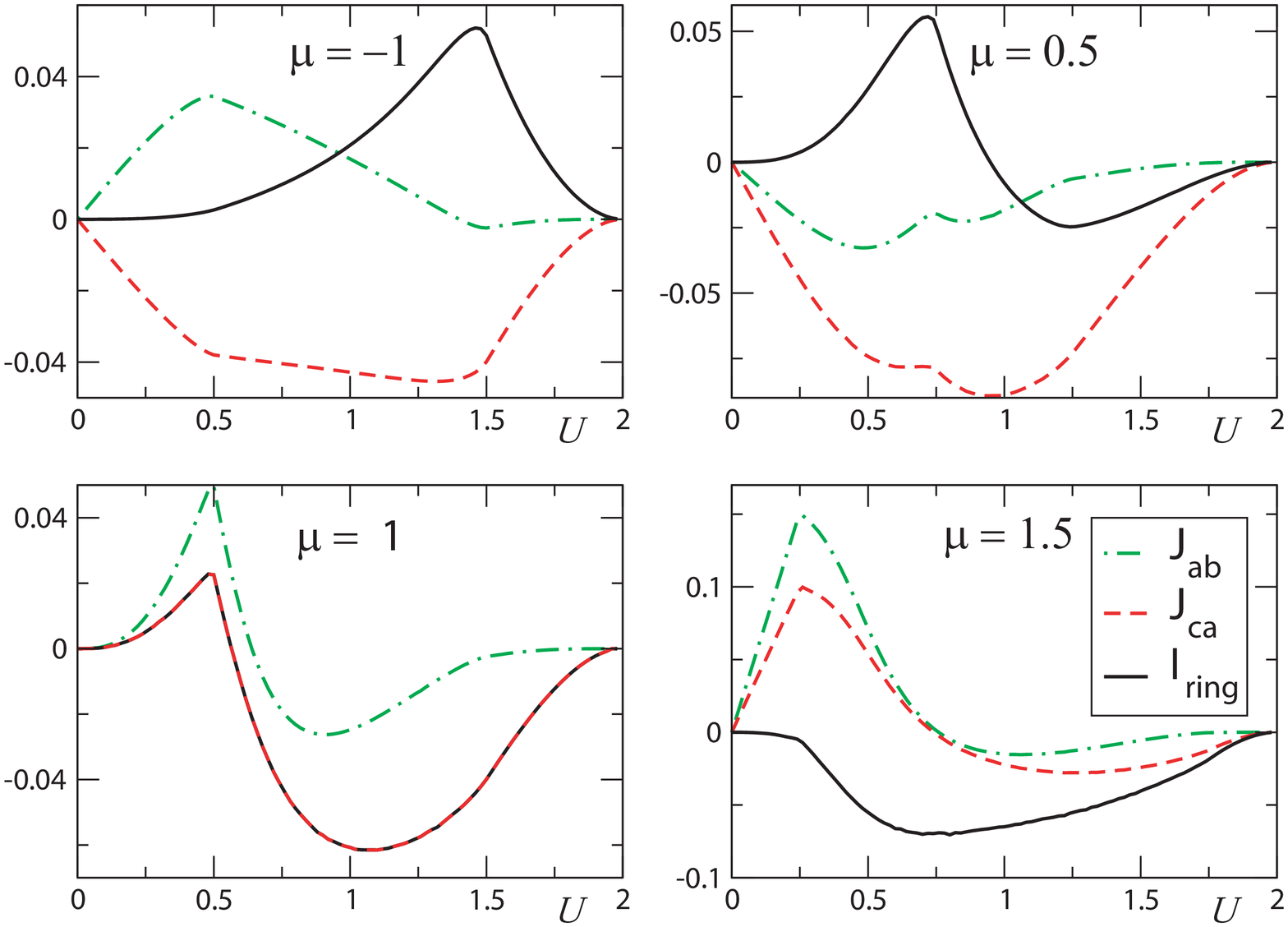}
\caption{(Color online) Plot of the bond currents $J_{ab}$ (green,
dot-dashed) and $J_{ca}$ (red, dashed) and of the
ring current $I_{\rm ring}$ (black, solid) defined in Eq. (\ref{iring2}) as a
function of the applied bias voltage $U_{R}=-U_{L}=U$ for different
values of the equilibrium chemical potential $\m$. Energies are in
units of the hopping parameter $t$ and currents are
in units of $tG_{0}$ with $G_{0}$ the quantum of conductance.}
\label{ivring}
\end{figure}

In Fig. \ref{ivring} we display the I/V characteristic of the ring
current as well as of the bond
currents along the $a-b$ and $c-a$ bonds for different values of the
chemical potential $\m=-1.5,0.-1.0,0.5,1.0$ in units of $t$.
The bond-currents are computed using a Landauer-like formula
derived in Ref. \onlinecite{noi}. In all
cases the ring current is quadratic in $U$ for small $U$ meaning that
the {\em ring conductance} is always zero.

 On the contrary, the bond currents are generally linear in $U$ except for special values of
$\m$ at which the {\em bond conductance} vanishes \cite{noi}. One of
these special values of $\m$ is $\m=1$. In this case the ring current
coincides with the bond current along the $c-a$ bond {\em for all}
values of $U$.
We also observe that $I_{\rm ring}$ has a maximum as a function of
$U$ and that the position of the maximum shifts towards high bias by
increasing the chemical potential. For negative $\m$ the maximum is
located in correspondence of the maximum of the bond currents while
for positive $\m$ is locate in correspondence of their minimum.
All currents correctly vanish for bias $U=2$ as the left continuum is
lifted by 2 while the right continuum is lowered by the same amount.
Since the
bandwidth is 4 the bias $U=2$ represents the minimum value of $U$ for
which there is no overlap between the left and right continua.

In a similar way one can compute $I_{\rm ring}$ for rings with $N$
sites, arms of different length and different hopping as well as
onsite energy parameters. We have verified that $I_{\rm ring}=0$ for rings
symmetrically connected, as physically expected.

In conclusion we have pointed out that for
quantum circuits containing closed loops the theory of quantum
transport needs to be extended in order to compute the loop
magnetic moments which couple to an external magnetic field. By a
suitable gedanken experiment we have been able to define a ring
current which is independent of the Peierls phase configuration
provided that $\a_{ab}+\a_{bc}+\a_{ca}=\f$ is kept constant. The
explicit calculation of $I_{\rm ring}$ in a ring connected to
one-dimensional
tight-binding leads show that $I_{\rm ring}$ is, in general, not
given by a linear combination of the bond currents, even though
there are common features. Our procedure differs from the one of
Ref. \cite{matthias} where the magnetic moment is obtained by an
average over the bond currents. We believe that our treatment can
be compared with experiment by measuring the ring magnetic
moments, and paves the way to include induction and self-induction
effects in quantum transport theory. In the extended theory, even
in the absence of an external magnetic field one will need to
consider a flux $\f=cLI_{\rm ring}$ where $L$ is the self-induction
coefficient.


\begin{thebibliography}{10}
\bibitem{units} In this work we will use atomic units and spin
indices will be omitted.
\end{thebibliography}
\end{document}